\documentclass[12pt]{article}
\usepackage{graphicx}
\usepackage{float}
\usepackage{graphicx,subfigure}
\usepackage{caption}
\usepackage[symbol]{footmisc}

\renewcommand{\thefootnote}{\fnsymbol{footnote}}

\newcommand\pubnumber{SNSN-323-63}
\newcommand\pubdate{\today}

\def\Title#1{\begin{center} {\Large #1 } \end{center}}
\def\Author#1{\begin{center}{ \sc #1} \end{center}}

\newcommand\pubblock{\rightline{\begin{tabular}{l} \pubnumber\\
         \pubdate  \end{tabular}}}
\newenvironment{Abstract}{\begin{quotation}  }{\end{quotation}}
\newenvironment{Presented}{\begin{quotation} \begin{center} 
             PRESENTED AT\end{center}\bigskip 
      \begin{center}\begin{large}}{\end{large}\end{center} \end{quotation}}
\def\Acknowledgements{\bigskip  \bigskip \begin{center} \begin{large}
             \bf ACKNOWLEDGEMENTS \end{large}\end{center}}

\def\beq{\begin{equation}}
\def\eeq#1{\label{#1}\end{equation}}
\def\eeqn{\end{equation}}

\def\beqa{\begin{eqnarray}}
\def\eeqa#1{\label{#1}\end{eqnarray}}
\def\eeqan{\end{eqnarray}}

\let\bar=\overbar

\def\Dslash{\not{\hbox{\kern-4pt $D$}}}
\def\dslash{\not{\hbox{\kern-2pt $\del$}}}

\def\msb{{\bar{\ssstyle M \kern -1pt S}}}

\begin{document}
\begin{titlepage}
\pubblock

\vfill
\Title{Muon Induced neutron measurement setup at TIFR}
\vfill
\Author{H. Krishnamoorthy$^{1,2}$, G. Gupta$^{3}$, A. Garai$^{1,2}$, A. Mazumdar$^{1,2}$,\\ A. Reza$^{3}$, S. Pal$^{4}$, V. Nanal$^{3}$, A. Shrivastava$^{2,5}$, and R.G. Pillay$^{3}$\footnote{presently at Department of Physics, Indian Institute of Technology Ropar, Rupnagar, Punjab, India 140001}}
$^{1}$India-based Neutrino Observatory, Tata Institute of Fundamental Research,\\ Mumbai - 400005, India\\
$^{2}$Homi Bhabha National Institute, Anushaktinagar, Mumbai - 400094, India \\
$^{3}$ Department of Nuclear \& Atomic Physics, Tata Institute of Fundamental Research, Mumbai - 400005, India\\
$^{4}$Pelletron Linac Facility, Tata Institute of Fundamental Research, Mumbai - 400005, India\\
$^{5}$Nuclear Physics Division, Bhabha Atomic Research Centre, Mumbai - 400085, India

\vfill
\begin{Abstract}
Cosmic muon induced neutrons in Pb are measured by direct neutron detection, using CLYC detectors. The detector set-up and preliminary results are presented.
\end{Abstract}
\vfill
\begin{Presented}
NuPhys2018, Prospects in Neutrino Physics\\
Cavendish Conference Centre, London, UK, December 19--21, 2018
\end{Presented}

\vfill
\end{titlepage}
\def\thefootnote{\fnsymbol{footnote}}
\setcounter{footnote}{0}

\section{Introduction}
Neutrons constitute an important background for neutrinoless double beta decay \cite{ndbd}, dark matter search, solar neutrino experiments and other rare event searches. Interaction of neutrons with nuclei can produce nuclear recoils in the detector active volume or lead to the production of excited nuclear states, 
leading to additional $\gamma$-ray background, often close to the region of the interest (ROI). Origin of the neutron background can be from the natural radioactivity (spontaneous fission and $\rm (\alpha,n)$ reactions) or from the interaction of cosmic ray muons with detector and surrounding materials. The secondary neutrons from the latter have energy range extending to several GeV and are relatively difficult to suppress with a prompt active veto or passive shielding methods \cite{meihime}.
Understanding the neutron background \cite{neha_jinst}, particularly in heavy targets such as Cu and Pb, which are often direct components of the detector or used as passive shield for $\gamma$-rays, is an important factor to achieve the required sensitivity levels. Previous measurements have reported results which are in disagreement with the Monte Carlo simulations \cite{minidex}. Study of muon induced neutron background in connection with neutrinoless double beta decay in $\rm ^{124}Sn$ ({\it TIN.TIN} experiment in India) \cite{vnanal} has been initiated at TIFR, Mumbai.  For this purpose a dedicated set up MINT (Muon Induced Neutron measurement setup at TIFR) has been made. This paper presents the details of MINT detector and preliminary results for neutron production in lead (Pb).

\section{MINT}

MINT is designed for direct detection of cosmic muon induced secondary neutrons. It employs novel, high efficiency CLYC $\rm (Cs_{2}LiYCl_{6}:Ce)$ detector \cite{clyc}, which can detect thermal and fast neutrons as well as $\gamma$-rays. It is compact and portable, to facilitate measurements at different locations and is presently installed at sea level. At present, MINT consists of two CLYC detectors, of size $\rm 1^{\prime\prime} dia \times 1 ^{\prime\prime} length$, surrounded by high density polyethylene (HDPE) layer of 10 cm thickness. The target material surrounds the CLYC detectors, which in turn is covered by plastic scintillators (see Figure~\ref{fig:mint}). 
The neutrons produced in the target material thermalize in HDPE and get captured via $\rm ^{6}Li (n,\alpha) ^{3}H$ reaction in the CLYC detector, with a unique light output at 3.2 MeVee. Excellent pulse shape discrimination (PSD) capabilities allow a clear separation for thermal neutrons and $\gamma$-rays. The thickness of HDPE was optimized  in an independent measurement with fast neutron source.  
The cosmic muons are detected using 
 plastic scintillators (50 cm $\times$ 50 cm $\times$ 1 cm).  Efficiency of plastic scintillators are measured to be $\sim$ 96\%. The data acquisition (DAQ) system consists of CAEN V1730B digitizer (500 MHz, 16 channel, 14 bit ADC, $\rm 2V_{pp}$)and data is recorded independently for each detector on an event by event basis  with time stamp, energy and PSD parameter. The coincidence search for correlated neutron events is performed offline with ROOT \cite{root} and C++ based algorithms.
 
\section{Measurements of neutron production in Pb}

\begin{figure}
\centering
\begin{minipage}{.5\textwidth}
  \centering
  \includegraphics[height=7.5 cm, width =7.1 cm,keepaspectratio]{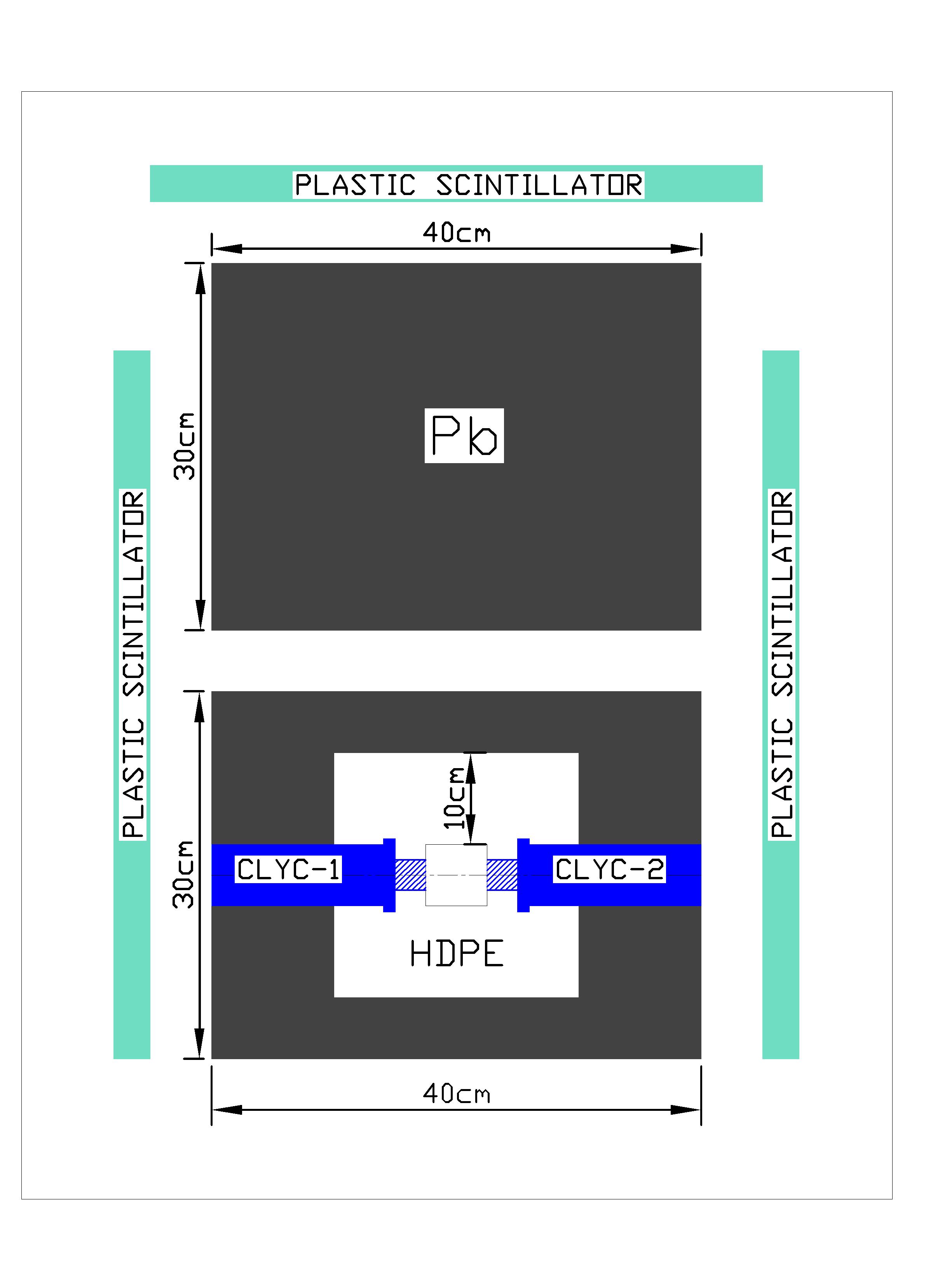}
  \captionof{figure}{Schematic of MINT set-up }
  \label{fig:mint}
\end{minipage}%
\begin{minipage}{.5\textwidth}
  \centering
  \includegraphics[height=7.5 cm, width =7.1 cm,keepaspectratio]{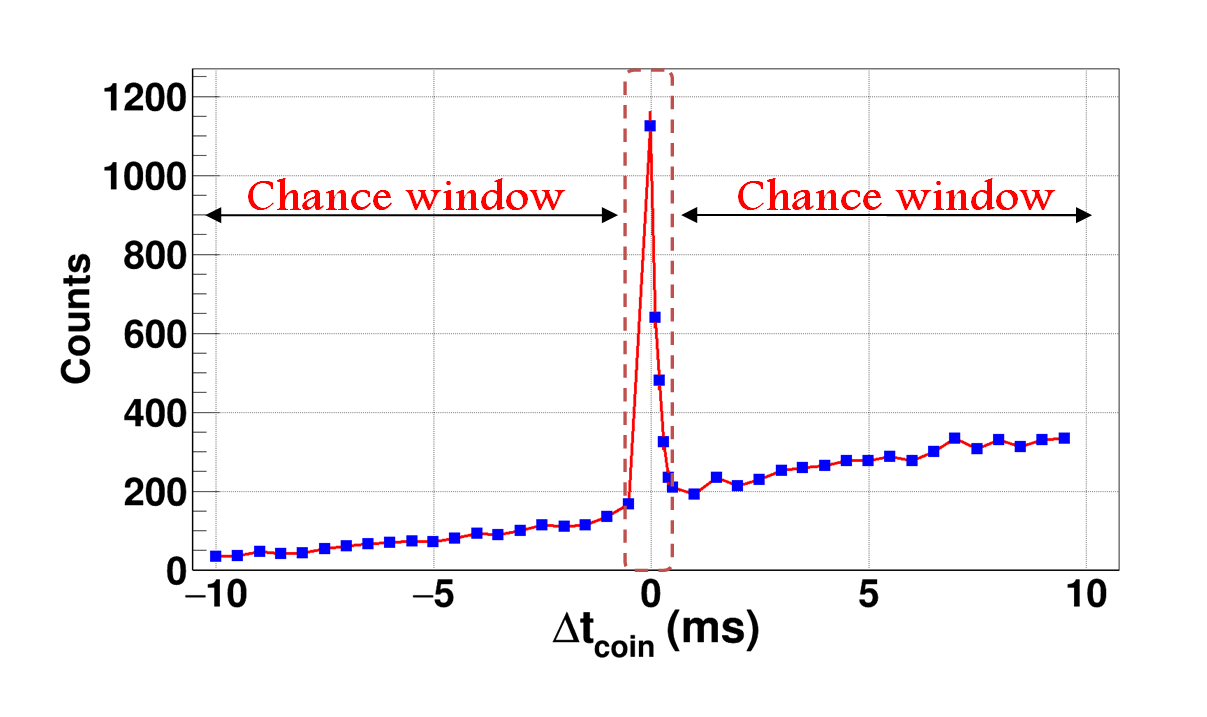}
  \captionof{figure}{$\rm T_{CLYC}-T_{plastic}$. Prompt window marked in dashed line. }
  \label{fig:tac}
\end{minipage}%
\end{figure}

 Since Pb is commonly employed shield material, the first set of measurements in MINT are performed with Pb.
 The Pb blocks of 30 cm height and  40 cm $\times$ 40 cm footprint ($\sim 758$ kg) are mounted as shown in the Figure \ref{fig:mint}. The choice of lead thickness was governed by the fact that
the chief contribution at sea level comes from stopping muons in Pb of $E_{\mu} <$ 500 MeV and 30 cm is comparable to range of $E_{\mu}\sim$ 500 MeV in Pb. The present data was taken for 40 days and focused on the detection of thermal neutrons. 
 Signature of an event is a coincidence between a muon signal in plastic scintillator and a subsequent delayed signal in the CLYC from thermal neutron capture. The coincidence was performed within a time window of $\rm T_{CLYC}-T_{plastic} $ = $\pm$ 10 ms. This window has been chosen to incorporate the neutron transport and thermalization time for MINT geometry, optimized from GEANT4 simulations~\cite{geant4}.


\begin{figure}
\centering     
\subfigure[Energy-PSD spectrum]{\label{fig:mint_2d}\includegraphics[height=7.5 cm, width =7.1 cm,keepaspectratio]{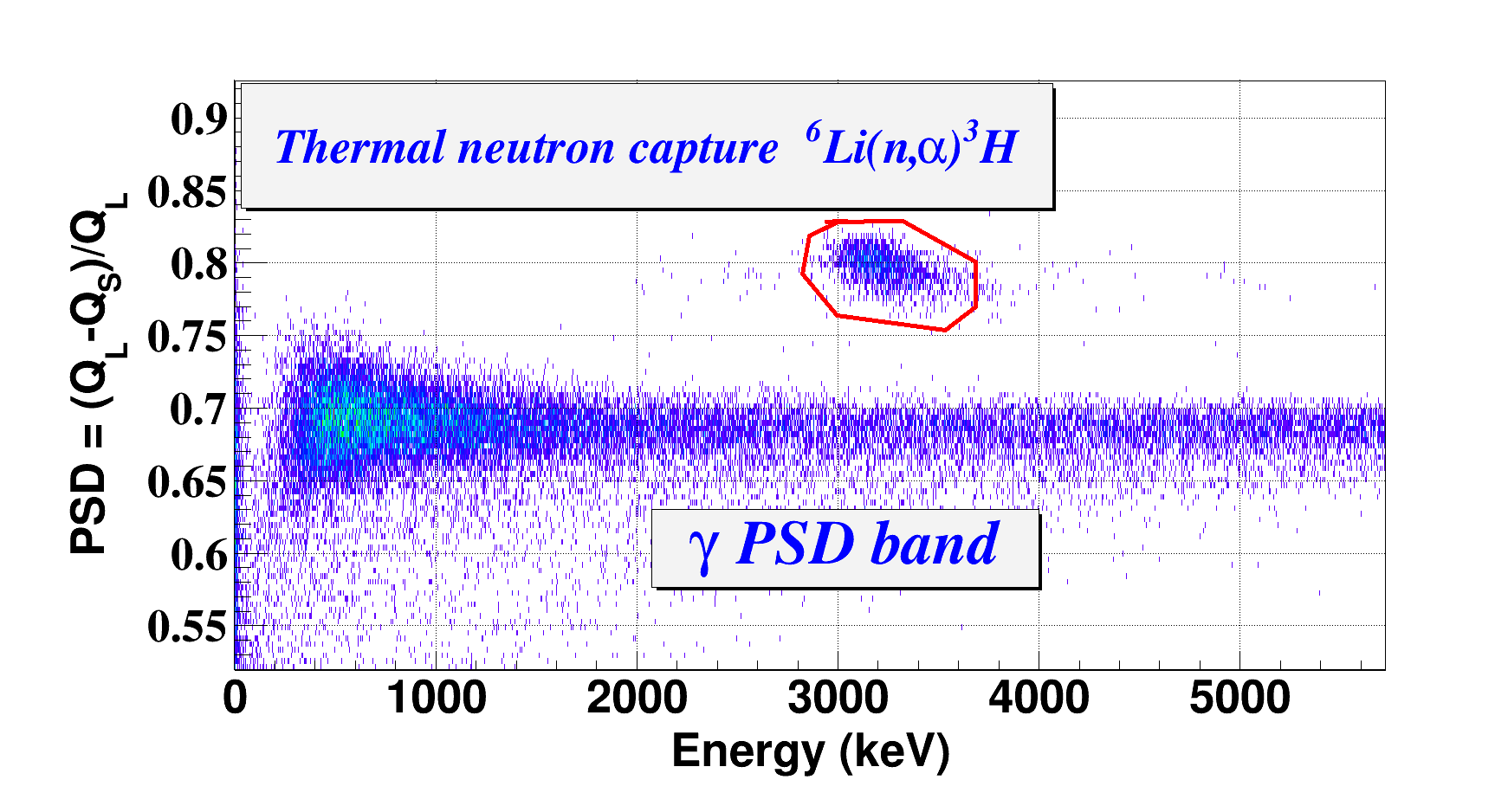}}
\subfigure[Projected counts in thermal neutron peak]{\label{fig:det_counts}\includegraphics[height=7.5 cm, width =7. cm,keepaspectratio]{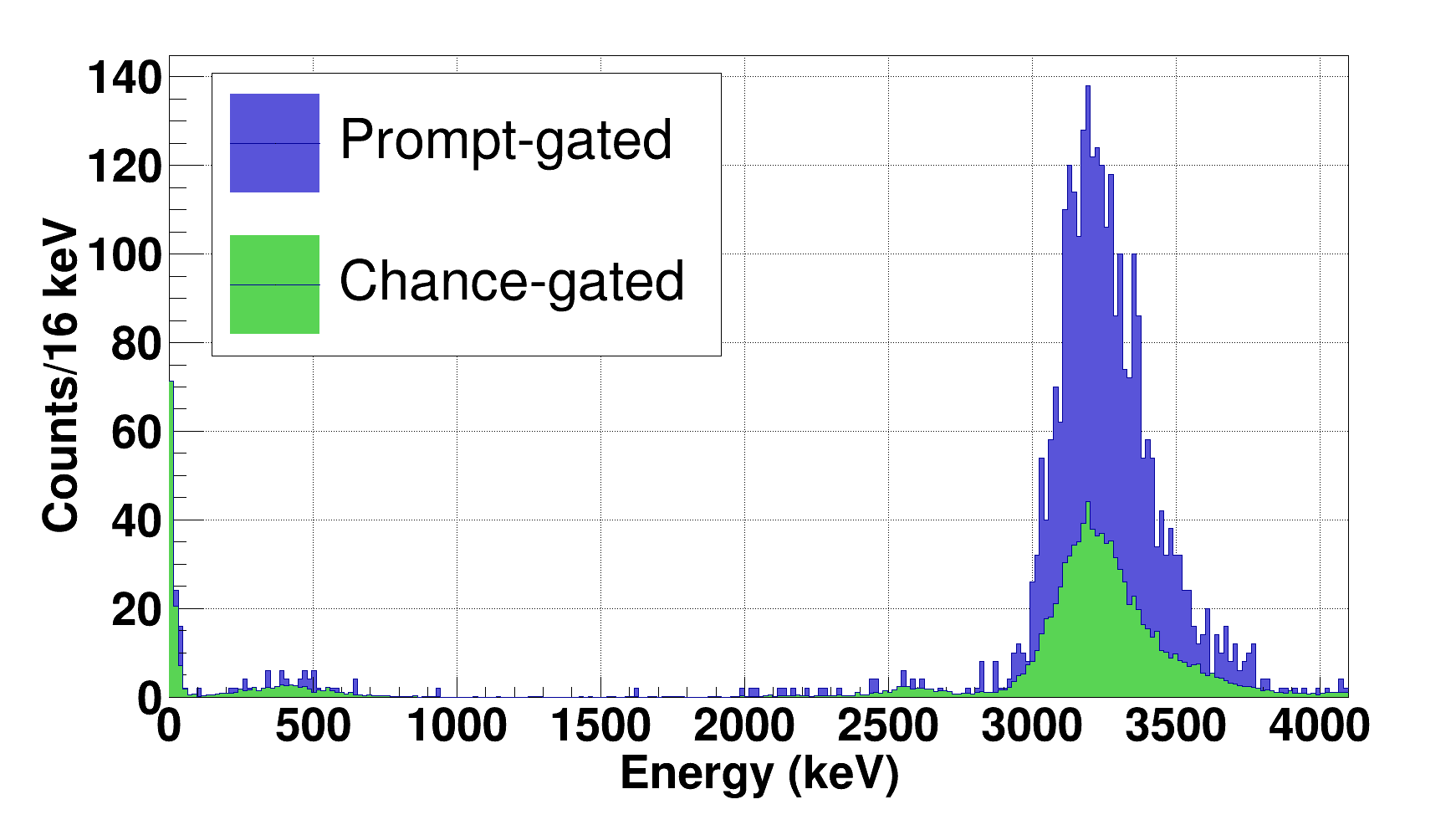}}
\caption{Prompt gated spectra in CLYC detector}
\end{figure}

Figure \ref{fig:tac} shows the time distribution between CLYC detectors and plastic scintillator. The time windows for prompt events is marked in the figure.  It can be seen that the background on left and right (pre and post muon interaction) are widely different and 
average of  both sides was used for chance correction. 
Figure \ref{fig:mint_2d} and 
Figure \ref{fig:det_counts} show the prompt-gated energy-PSD and  energy spectra, respectively.
\begin{table}[H]
\centering
\caption{\label{tab:i}Measured yield of neutrons in MINT. $\rm (T_{data}$ = 40 days )}
\smallskip
\begin{tabular}{|c|c|}
\hline
 & counts\\
\hline

$\rm T_{prompt}$  & 2802(306)\\
$\rm T\footnotemark_{chance}$  & 869 (239)\\
Chance corrected counts/day & 48(14)\\

\hline
\end{tabular}
\end{table}

\footnotetext{scaled to prompt time width}

The errors bars in the counts are statistical. Independent measurements with $\rm ^{252}Cf$ neutron source ($\rm E_{n} \sim 10 MeV$) and CLYC detector were carried out to estimate the fractions of neutrons thermalized and transmitted for a given  HDPE thickness. Combining this with intrinsic efficiency of CLYC, measured to be 25\% \cite{clyc_dae}, fast neutron production from cosmic muons is estimated from above data as  8.1 $\pm$2.4 neutrons/g$\rm cm^{-2}$/day, for an effective Pb thickness of 30 cm. Monte Carlo simulations for MINT is in progress.

\section{Summary}
 A dedicated set-up MINT, for measuring muon induced neutrons in different materials has been set up at TIFR with novel CLYC detectors. Preliminary measurements with Pb target have been carried out and the neutron yield from cosmic muons is measured to be 8.1 $\pm$2.4 neutrons/g$\rm cm^{-2}$/day. It is also proposed to investigate the fast neutron and $\gamma$-ray detection (from $\rm ^{1}H(n,\gamma)^{2}H$ in HDPE) for a complete study of cosmic muon induced neutrons.
\pagebreak

\Acknowledgements
We thank Mr. K.V. Divekar, Mr. M.S. Pose, Mr. S. Mallikarjunachary for the assistance with the measurements. The support of \textit{TIN.TIN} and INO collaborations is gratefully acknowledged.

\end{document}